\RequirePackage{amsmath}
\documentclass[runningheads]{llncs}

\usepackage{varioref}
\usepackage{amsmath,amssymb,amsfonts}
\usepackage{algorithmic}
\usepackage{graphicx}
\usepackage{textcomp}
\usepackage{xcolor}
\usepackage{colortbl}
\usepackage[switch]{lineno}
\usepackage{url,hyperref}
\usepackage{adjustbox}
\usepackage{float}
\usepackage[T1]{fontenc}
\usepackage[utf8]{inputenc}
\usepackage{tikz}
\usepackage{pifont}
\usepackage{scalerel}
\usepackage{ifthen}
\usepackage{pdfpages}
\usepackage{algorithmic}
\usepackage{graphicx}
\usepackage{textcomp}
\usepackage[switch]{lineno}
\usepackage{url}
\usepackage{adjustbox}
\usepackage{float}
\usepackage[T1]{fontenc}
\usepackage{pdfpages}
\usepackage{tikz}
\usepackage{pifont}
\usepackage{scalerel}
\usepackage{ifthen}
\usepackage{tabularx}
\usepackage{subcaption}
\usepackage{caption}
\usepackage{wasysym}
\usepackage{marvosym}
\usepackage{booktabs}
\usepackage{multicol}
\usepackage{multirow}
\usepackage{ragged2e}
\usepackage{tabularx}

\usepackage{nicematrix,tikz}
\usepackage{pgfplots}
\usepackage{makecell}
\usepackage{tocbibind}

\usepgfplotslibrary{fillbetween}
\usetikzlibrary{patterns}
\usetikzlibrary{math}
\usepgfplotslibrary{groupplots}
\pgfplotsset{compat=newest}
\usepackage{float}

\usepackage{bbding}

\usepackage[linesnumbered,ruled,vlined]{algorithm2e}

\usepackage{academicons}
\usepackage{xcolor}

\usepackage{fancyhdr} 
\usepackage{lastpage}
\def\BibTeX{{\rm B\kern-.05em{\sc i\kern-.025em b}\kern-.08em
    T\kern-.1667em\lower.7ex\hbox{E}\kern-.125emX}}

\SetKwInput{KwInput}{Input}                
\SetKwInput{KwOutput}{Output}              
\SetKwInput{KwFunction}{Function}              

\definecolor{Gray}{gray}{0.85}
\definecolor{LightCyan}{rgb}{0.88,1,1}

\usepackage[natbib=true, style=numeric,sorting=none]{biblatex}
\begin{document}

\title{PriCE: Privacy-Preserving and Cost-Effective Scheduling for Parallelizing the Large Medical Image Processing Workflow over Hybrid Clouds}



\author{
    Yuandou Wang\inst{1} \href{https://orcid.org/0000-0003-4694-9572}{\includegraphics[scale=0.01]{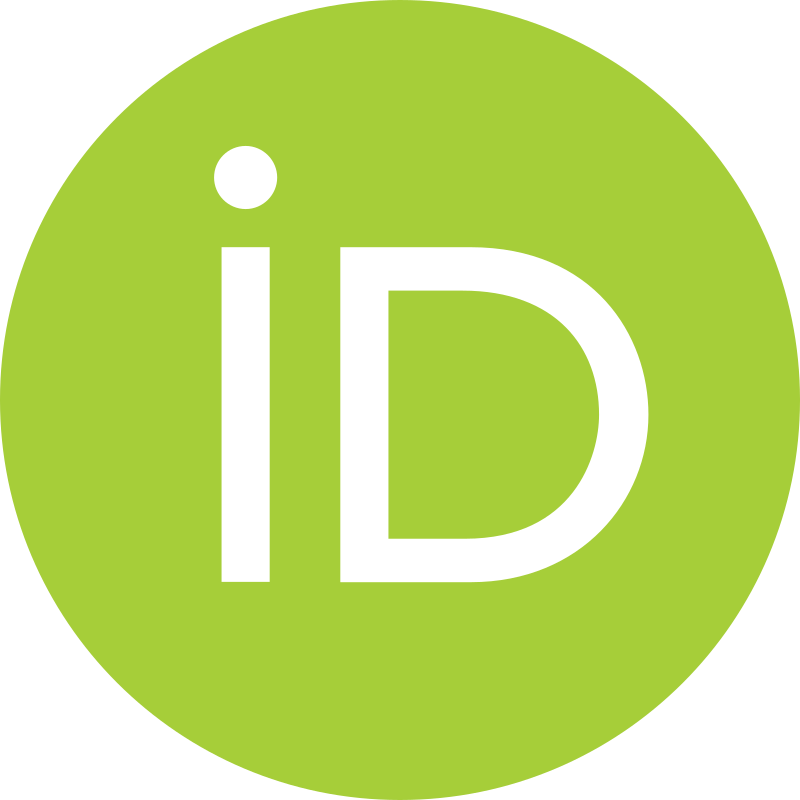}} \and
    Neel Kanwal\inst{2} \href{https://orcid.org/0000-0002-8115-0558}{\includegraphics[scale=0.01]{figs/orcid.png}} \and 
    Kjersti Engan\inst{2} \href{https://orcid.org/0000-0002-8970-0067}{\includegraphics[scale=0.01]{figs/orcid.png}} \and
    Chunming Rong\inst{2} \href{https://orcid.org/0000-0002-8347-0539}{\includegraphics[scale=0.01]{figs/orcid.png}} \and
    Paola Grosso\inst{1} \href{https://orcid.org/0000-0003-4600-9812}{\includegraphics[scale=0.01]{figs/orcid.png}} \and
    Zhiming Zhao\inst{1} \href{https://orcid.org/0000-0002-6717-9418}{\includegraphics[scale=0.01]{figs/orcid.png}}
}

\authorrunning{Y. Wang et al.}
%
\institute{Multiscale Networked Systems, University of Amsterdam, The Netherlands 
\email{\{y.wang, p.grosso, z.zhao\}@uva.nl} \and
Department of Electrical Engineering and Computer Science, University of Stavanger, Norway \\
\email{\{neel.kanwal, kjersti.engan, chunming.rong\}@uis.no} 
}
\maketitle              
\begin{abstract}
Running deep neural networks for large medical images is a resource-hungry and time-consuming task with centralized computing. Outsourcing such medical image processing tasks to hybrid clouds has benefits, such as a significant reduction of execution time and monetary cost. However, due to privacy concerns, it is still challenging to process sensitive medical images over clouds, which would hinder their deployment in many real-world applications. To overcome this, we first formulate the overall optimization objectives of the privacy-preserving distributed system model, i.e., minimizing the amount of information about the private data learned by the adversaries throughout the process, reducing the maximum execution time and cost under the user budget constraint. We propose a novel privacy-preserving and cost-effective method called PriCE to solve this multi-objective optimization problem. We performed extensive simulation experiments for artifact detection tasks on medical images using an ensemble of five deep convolutional neural network inferences as the workflow task. Experimental results show that PriCE successfully splits a wide range of input gigapixel medical images with graph-coloring-based strategies, yielding desired output utility and lowering the privacy risk, makespan, and monetary cost under user's budget. 

\end{abstract}

\section{Introduction}\label{sec:intro}
Modern medical image processing techniques utilize deep neural networks to extract hidden patterns and make predictions; however, running such machine learning-based inferences for large medical images is resource-hungry and time-consuming when computing resources are limited. Cloud computing can provide ample and highly scalable storage, computational resources, and ubiquitous access for distributed processing tasks. Hybrid Clouds (HCs) combine the economies and efficiencies of public cloud with the security and control of private cloud~\cite{mazhelis2012economic}. However, privacy concerns significantly complicate the development of an optimal cloud resource allocation plan for outsourcing computations on sensitive data processing tasks: (1) medical images often contain privacy-sensitive information in their metadata, which cannot be directly outsourced to the public cloud due to the risk of data leakage, (2) assigning distributed processing workloads to available cloud resources to meet multiple user requirements such as the reduction of time and monetary cost, known as Multi-Objective Optimal (MOO) workflow scheduling, is a typical NP-hard problem~\cite{sharif2016privacy}.


The problem of privacy-preserving and cost-effective scheduling in large (e.g., gigapixel) medical image processing over HCs has not yet been studied in detail, and we observed that this problem exhibits unique characteristics. For instance, although patching the original large image in a grid and dividing the patch-level dataset into multiple sub-datasets are common practices, the strategy employed for these practices is critical for both privacy preservation and resource provisioning for deployment in the cloud. Furthermore, although several studies have addressed workflow privacy in the context of cloud technology~\cite{sharif2016privacy, lei2022privacy}, there is a need for precise measurements of privacy metrics that align with the specific privacy-preserving approaches used for workflow scheduling. 

This work aims to overcome those limits and solve the problem of privacy-preserving and cost-effective distributed inference tasks over clouds. To address this, we first formulate the research problem that minimizes the vulnerability, reduces the monetary cost, and minimizes the maximum execution time of the privacy-preserving distributed system under constraints. We propose a novel privacy-preserving and cost-effective algorithm called PriCE. Our experimental evaluation reveals the benefits of PriCE in privacy-preserving and cost-effective workflow scheduling when answering the following three sub-research questions: (I) ``What are the trade-offs among privacy, cost, and execution time as split strategies related to the distributed processing of large images?'' (II) ``Can privacy-preserving split strategies improve resource planning and lead to Pareto optimality?'' and (III) ``How different are the Pareto optimal solutions from different split strategies?''

Our main contributions can be summarized as follows:
\begin{itemize}

    \item We design and implement PriCE, which consists of multiple image-splitting strategies, image label perturbation, and multi-objective optimization procedures that can seek the Pareto front of resource provisioning for the privacy-preserving and cost-effective system model. 

    \item We demonstrate how to analyze, quantify, compare, and understand different split strategies within PriCE and obtain the final assignment, estimated objective values of the cloud instances, by conducting experiments based on a use case for artifact detection tasks on gigapixel medical images. 
    
\end{itemize}

The remainder of this paper is structured as follows. Section~\ref{sec:related} presents related work to privacy-aware workflow scheduling in HCs. In Section~\ref{sec:system}, we propose our system model, provide critical metrics used for the methodology and evaluation, and formulate the research problem. Section~\ref{sec:method} illustrates our proposed approach for problem-solving and Section~\ref{sec:experiments} details the experimentation and evaluation. Finally, we conclude our work in Section~\ref{sec:conclusion}.

\section{Related Work}\label{sec:related}

The scheduling of privacy-aware workflows has garnered increased attention in recent years, especially aiming to minimize costs and processing time while ensuring compliance with privacy requirements. 
Sharif \textit{et al.}~\cite{sharif2016privacy} target a resource allocation map based on privacy privileges over HCs that combine private, community, and public clouds, while using a healthcare workflow that consists of private, semi-private, and public tasks as a case for problem modeling. Their objective is to minimize the overall execution cost of workflows while satisfying concerns about their privacy and deadline. Zhou \textit{et al.}~\cite{zhou2019privacy} study the European (EU) data protection regulation --- GDPR\footnote{General Protection Data Regulation (GPDR). \url{https://gdpr.eu/}} in the geo-distributed cloud and formulate the geo-distributed process mapping problem that minimizes the cost of workflow applications while meeting data privacy constraints regarding the restrictions of the data movement between different cloud data centers. Lei \textit{et al}.~\cite{lei2022privacy} define a deadline-constrained cost optimization problem in a HC under the deadline and privacy constraints. Similarly, Wen \textit{et al.}~\cite{wen2020scheduling} propose a multi-objective privacy-aware scheduling algorithm to obtain a set of Pareto trade-off solutions between execution makespan and cost reductions while meeting the set of privacy protection constraints. As presented in Table~\ref{tab:literature}, these works are close to our research problem in terms of data and task privacy constraints, time performance improvement, and cost reduction for resources planning over HCs; nonetheless, our work significantly differs from their works. 
\begin{table}[!thb]
\scriptsize
    \centering
    \caption{Comparisons of the problem formulation. }
    \label{tab:literature}
    {\renewcommand{\arraystretch}{1.5}%
    \resizebox{0.98\textwidth}{!}{
    \begin{tabular}{lcccc}
    \toprule
    \multirow{2}{*}{ {Problem Model}} & \multicolumn{3}{c}{ {Optimization Objective}} & \multirow{2}{*}{ {Constraint(s)}} \\ \cline{2-4}
    & \multicolumn{1}{c}{ {Time}} & \multicolumn{1}{c}{ {Cost}} & \multicolumn{1}{c}{ {Privacy}} &\\ 
    \midrule
     {Sharif \textit{et al.}~\cite{sharif2016privacy}} &  {\color{red}\XSolidBrush} &  {\color{green}\CheckmarkBold} &  {\color{red}\XSolidBrush} &  {task/data privacy, deadline} \\
     {Lei \textit{et al}.~\cite{lei2022privacy}} &  {\color{red}\XSolidBrush} &  {\color{green}\CheckmarkBold} &  {\color{red}\XSolidBrush} &  {deadline, privacy} \\
     {Zhou \textit{et al.}~\cite{zhou2019privacy}} &  {\color{red}\XSolidBrush} &  {\color{green}\CheckmarkBold} &  {\color{red}\XSolidBrush} &  {data privacy, GDPR, deadline}\\
     {Wen \textit{et al.}~\cite{wen2020scheduling}} &  {\color{green}\CheckmarkBold} &   {\color{green}\CheckmarkBold} &  {\color{red}\XSolidBrush} &  {privacy protection}\\
     {\textbf{Ours}} &   {\color{green}\CheckmarkBold} &  {\color{green}\CheckmarkBold} &  \color{green}\CheckmarkBold & {user's budget} \\
    \bottomrule
    \end{tabular}
   } 
   }
\end{table}

We study the scheduling problem that minimizes the quantified amount of information about private data learned by the adversaries, lowers financial cost, and reduces the maximum execution time with different data-split strategies. To the best of our knowledge, this is a unique problem statement since the majority of the related works leave privacy as a constraint, instead of a quantified optimization objective.

\section{Problem Formulation}\label{sec:system}

\subsection{System Model}
\begin{figure*}[!htb]
    \centering
    \includegraphics[width=0.98\textwidth]{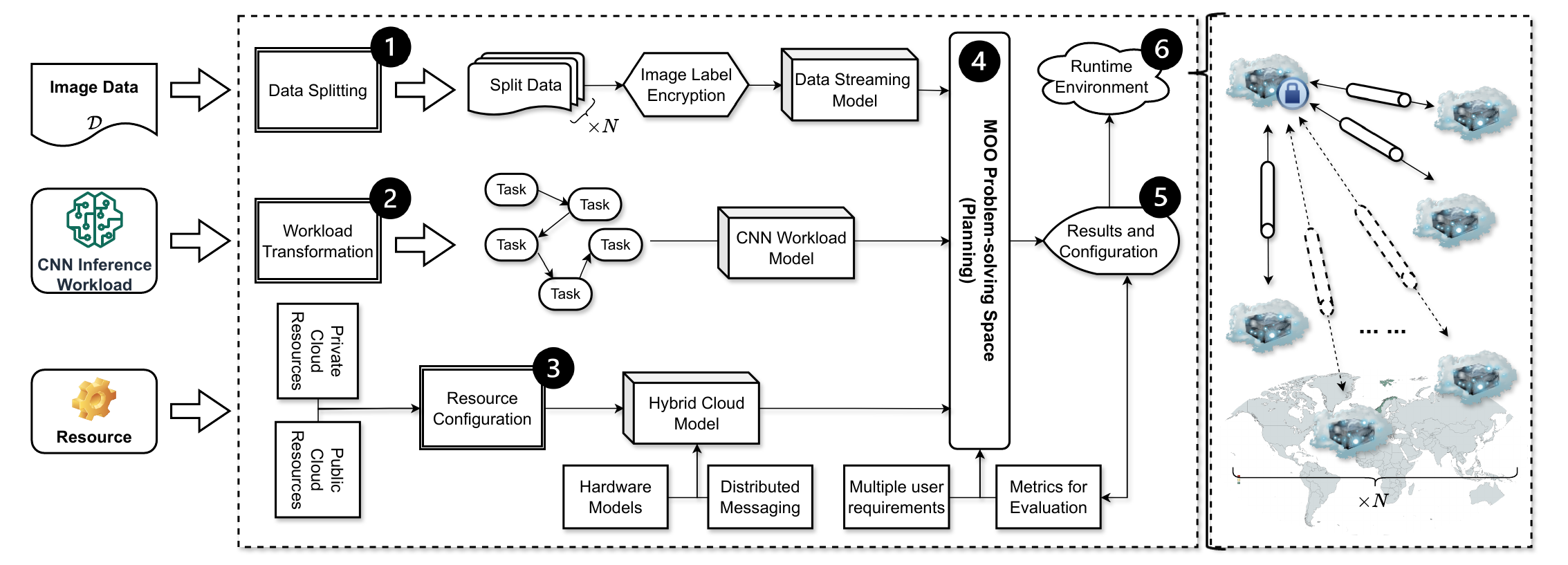}%
    \caption{Workflow of the system model and its related application scenario.}
    \label{fig: distributed data processing}
\end{figure*}

We consider a system as depicted in Fig.~\ref{fig: distributed data processing}.
We remove the privacy-sensitive metadata from the original image $\mathcal{D}$ before splitting the large image into many image tiles to introduce data parallelism. We apply a privacy-preserving data splitting procedure in step~\ding{182}, in conjunction with image label encryption and image object serializing to preserve the data privacy. Data splitting aims to protect data privacy by fragmenting sensitive data and storing the fragments in different locations so that individual parts do not disclose identities or confidential information~\cite{domingo2019privacy}. In step~\ding{183}, we transform the Convolutional Neural Network (CNN) inference model into several reusable fine-grained computational tasks; configuring the available resources such as private and public cloud resources in step~\ding{184} can facilitate making plans for resource provision. 

Based on the prepared data, workload, and HC model, step~\ding{185} maps available resources to various workloads in a manner that optimizes their utilization and satisfies user requirements. The privacy-preserving and cost-effective problem is a classic MOO scheduling problem, where $N$ workloads and split privacy-preserved datasets have to be scheduled on $M$ identical cloud instances over HCs, with the multi-objective functions that 
minimize the information about sensitive data learned by adversaries, the financial expense of the total instances consumed, and the maximum execution time of the instance that completes the last workload, i.e., makespan. 
After obtaining the estimated results and configuration in step~\ding{186}, the application will be deployed and executed onto the runtime environment. The system should ensure that data storage and workload execution remain in place and continue to be effective even among changes, such as downtime, errors, or attackers, to the system or emerging threats, according to step~\ding{187}~\cite{wang2023towards}. Note that when constructing this model, we refer to the prior research experience and user requirements from several EU projects such as CLARIFY\footnote{CLARIFY project. \url{http://www.clarify-project.eu/}}, BlueCloud2026\footnote{Blue-Cloud2026 project. \url{https://blue-cloud.org/about-blue-cloud-2026}}, ENVRI-Hub NEXT\footnote{ENVRI-Hub NEXT. \url{https://envri-hub.envri.eu/}}, and LifeWatch ERIC\footnote{LifeWatch ERIC. \url{https://www.lifewatch.eu/}}. 

\subsection{Privacy-Preserving and Cost-Effective Metrics}
To satisfy user demands, privacy-preserving and cost-effective data processing should consider goals such as privacy preservation, makespan, and monetary cost. Additionally, we use the semi-honest threat model to evaluate the robustness of the system~\cite{domingo2019privacy}. HC service providers honestly fulfill their role in the storage and processing tasks, but may inspect the information that users store or process. The attacker typically tries to collude with a number of processing nodes storing the datasets to infer and reason sensitive information of the other nodes to gain insights without directly altering the data or system. We assume that once the attacker obtains the image datasets stored on the nodes, he will try to reconstruct the original image data by using sensitive information from image labels, e.g., the coordinates $(x, y)_\text{coord}$. 
The system should evaluate the results and configuration of the MOO problem-solving before the sub-datasets and CNN inference models are assigned to planned infrastructures.

To quantify the privacy-preserving goals, we introduce information-theoretic metrics since they assume a stronger adversary and are more efficient concerning both communication and computational demands~\cite{kraskov2004estimating, li2021privacy}. Let $\mathcal{N}$ be the number of datasets $\{d_{p,1}, d_{p,2},..., d_{p,N}\}$ split from the entire image $\mathcal{D}$; the set $\{d_{e,1}, d_{e,2},..., d_{e,N}\}$ denotes the encrypted datasets. Let $Z_i$ and $S_i$ be the encrypted and private information of $d_{e,i}$ and $d_{p,i}$ ($i\in \mathcal{N}$), respectively. We denote the entropy that measures the randomness of $S_i$ as $H(S_i) = -\sum^{n}_{j=1}p_jlog_2p_j$, where $p_j$ denotes the probability of the unique coordinate information in $S_i$. The mutual information $I(S;Z)=H(S)-H(S|Z)$ between two random variables $S$ and $Z$ measures the dependence between $S$ and $Z$, quantifying the average reduction in uncertainty about $S$ that results from learning the value of $Z$.

\paragraph{\textbf{Output Utility.}}
The output utility is to measure how close the estimated value $\hat{Y}_i$ of a privacy-preserving distributed processing algorithm is to its desired output $Y_i$, for each node $i \in \mathcal{N}\subset M$,
\begin{equation}\label{eq:utility}
    u_i = I(Y_i;\hat{Y}_i), \quad \forall i \in \mathcal{N}
\end{equation}
where $0\leq u_i\leq I(Y_i; Y_i)$ and $u_i=I(Y_i, Y_i)$ indicates perfect output utility~\cite{li2021privacy}.  

\paragraph{\textbf{Privacy Risk.}}
Let $\mathcal{V}$ denote the set of random variables containing all information collected by the adversaries throughout the whole process. 
The individual privacy of honest node $i\in \mathcal{N}_h$ quantifies the amount of information about the private data $S_i$ learned by the adversaries, which is given by, 
\begin{equation}
    \rho_i = I(S_i; \mathcal{V}), \quad \forall i \in \mathcal{N}_h \subset \mathcal{N}
\end{equation}
The smaller $\rho_i$, the more private the data is, hence, the lower the privacy risk the data is. Based on the definitions of the adversary model, the lower bound on individual privacy risk is formally stated by, 
\begin{equation}
    \rho_{i, min} = I(S_i; \{S_j, \hat{Y}_j\}_{j\in \mathcal{N}_c}), \quad \mathcal{N}_c = \mathcal{N}-1\label{eq:rho_i_min}
\end{equation}
where adversaries always have knowledge of the private data $\{s_j\}_j = S_j$ and estimated outputs over corrupted nodes $\{\hat{y}_j\}_j = \hat{Y}_j$, in which the maximum number of corrupted nodes $\mathcal{N}_c$ denotes $\mathcal{N}-1$ out of $\mathcal{N}$~\cite{li2021privacy}. 

\paragraph{\textbf{Total Cost and Makespan.}} To measure the total cost and makespan of employing cloud instances $\mathcal{N}$ over the HC model, we consider the pay-as-you-go pricing model. Let $p_k$ be the unit price of the $k^{th}$ cloud instance over the HC that consists of commercial and private cloud resources $M$. The total monetary cost is given by, 
\begin{equation}
    \text{Cost} = \sum_{k=1}^{\mathcal{N}} (T^{(\text{compt.})}_{k}+T^{(\text{comm.})}_{k})\times p_k \times x_k, \quad k\in M 
\end{equation}
where $x_k\in \vec{x}$ is a boolean value that if sub-dataset $d_{k}$ from $\mathcal{D}$ and the inference model are mapping to the cloud instance $k$, then $x_k$ equals 1; otherwise, 0. If $k$ belongs to commercial cloud instances, then $p_k\in R^{+}$; otherwise, $p_k=0$. $T^{(\text{comm.})}_{k}$ is the communication time when transferring the inference model and encrypted sub-dataset $d_{e,k}$ to the cloud instance $k$ and $T^{(\text{compt.})}_{k}$ is the computation time when running the workload on $k$, respectively. Similarly, the makespan over distributed processing nodes is given by, 
\begin{equation}
    \text{Makespan} = \max((T^{(\text{compt.})}_{k}+T^{(\text{comm.})}_{k}) \times x_k), \quad k\in M 
\end{equation}
where the compute capacity, geographical location, and network bandwidth of the instance $k$ impact the time of executing the privacy-preserving distributed application that typically consists of the time cost of on-site computation and communication overhead.

\subsection{Multi-Objective Optimization} \label{sec: problem_formulation}
The overall optimization objective of the system model intends to minimize the average lower bound on privacy risk $f_1$, the total monetary cost $f_2$, and the maximum completion time $f_3$ over a HC:

\begin{align}
    \min f_1 = & \text{Average minimal privacy risk} = \bar{\rho}_{min}(s), \quad \forall s \in \mathcal{S} \label{eq:f1}\\
    \min f_2 = & \text{Cost} = \sum_{k=1}^{\mathcal{N}} (T^{(\text{compt.})}_{k}+T^{(\text{comm.})}_{k})\times p_k \times x_k, \quad \forall k \in M \label{eq:f2}\\
    \min f_3 =& \text{Makespan} = \max((T^{(\text{compt.})}_{k}+T^{(\text{comm.})}_{k}) \times x_k), \quad \forall k \in M \label{eq:f3}
\end{align}
with regards to the following constraints:

--- The average privacy risk $\bar{\rho}_{min}$ is sensitive to privacy-preserving data-splitting strategy $s$ in the set $\mathcal{S}$, since each strategy $s$ generates unique sub-datasets of different sizes and image labels. 

--- The total number of split datasets $\mathcal{N}$ is expected to be the minimal number of the employed distributed processing nodes, which is limited to the maximum available resources of the HC model $M$.

--- Each split dataset and its corresponding inferences should be mapped to only one instance at a time. Meanwhile, one instance only runs one processing task per time. The total monetary cost $f_2$ is limited by the user's \textit{budget}. 



\section{PriCE: Privacy-Preserving and Cost-Effective Solution}\label{sec:method} 
This section presents the proposed PriCE solution and its algorithm pseudocode, as detailed in Algorithm \ref{alg: privacy method}. The solution primarily involves two key components: (1) privacy-preserving image splitting using graph-coloring, and (2) 3D Pareto trade-off solutions for resource planning.
\begin{algorithm}[!thb]
\DontPrintSemicolon
  \KwInput{The original image $\mathcal{D}$, patch size $p$, split strategies $\mathcal{S}$, inference workloads $\Theta$, and cloud instances $M$, budget $B$}
  \KwOutput{A new label system $A_e$, Pareto optimal solutions $\vec{x}^*$, and encrypted split datasets $d_{e,1},d_{e,2},..., d_{e,N}$.}

  Patch set $D \gets$ create\_patches($\mathcal{D}$, $p$)

  $G \gets$ abstract\_a\_graph($D$) 

  $\mathcal{N}, C \gets \textbf{graph\_coloring}(G, s)_{s\in \mathcal{S}}$
  
  $d_{p,1}, ...,d_{p,N} \gets$ divide $D_p$ into $\mathcal{N}$ sub data sets

  \For{$d_{p,k} \in d_{p,1}, ...,d_{p,N}$}{
    Data matrix $A_{p}$ of size ($a \times b$) $ \gets S_k \gets d_{p,k} $
    
  $\bar{x} \gets$ mean($A_{p}$) 
  
  $\sigma \gets$ std($A_{p}$) 
  
  $A_{p,c} \gets (A_p - \bar{x})/\sigma $ \tcp*{Normalize the data }

  $\text{\textbf{Cov}}(A_{p,c}) \gets \frac{1}{a-1} A_{p,c}^T A_{p,c}$ \tcp*{Compute the covariance matrix of $A_{p,c}$}

  $\lambda, \vec{V} \gets$ \textbf{eig}($\text{\textbf{Cov}}(A_{p,c})$) \tcp*{Compute eigenvalues and eigenvectors}
  
  $\lambda_{sort}, \vec{V}_{sort} \gets$ \textbf{sort\_eig}($\lambda$, $\vec{V}$)

  $\vec{V}_k \gets$ first $k$ columns of $\vec{V}_{sort}$ 

  $A_{e} \gets A_{p,c} \vec{V}_k$ \tcp*{Transform the original coordinate data}
  
  $d_{e,k} \gets \text{rename}(d_{p,k}, A_{e}, r)$ \tcp*{Transform labels with random values $r$ and perturbed coordinates $A_{e}$}
  }

  $\bar{\rho}_{min}(s)_{s\in \mathcal{S}} \gets \rho_{i, min}(s)_{s\in \mathcal{S}}, i\in \mathcal{N} \gets \textbf{Eq.~\ref{eq:rho_i_min}}$ 
  
  $totalC, maxTime, \vec{x} \gets \textbf{prob.solve}(\mathcal{N}, \Theta, M, B)$, derived from \textbf{Eqs.}~\ref{eq:f1},~\ref{eq:f2}, and~\ref{eq:f3} 
    
  $\vec{x}^* \gets \text{plot\_Pareto\_3D}(\vec{x})$ \tcp*{plot the 3D Pareto trade-off solutions}
  
    \Return $\vec{x}^*, A_e, \{d_{e,1},d_{e,2},..., d_{e,N}\}$

    \caption{PriCE Method\CheckedBox}\label{alg: privacy method}
\end{algorithm}

\subsection{Privacy-Preserving Image Splitting with Graph-Coloring} \label{privacy mechanism design}
To cope with the diverse image samples of the privacy-preserving data-splitting procedure, we abstract the entire image as a grid graph where different patches with pixel size $p\times p$ are cropped from the original image $\mathcal{D}$, containing sensitive image labels and objects. The image label contains sensitive coordinate information to reconstruct the image and guide the outcome. One essential hypothesis is that the more adjacent image patches an attacker obtains, the higher the probability he could succeed in restoring the entire original image. 

Let $G=(V,E)$ be a graph extracted from the entire patch dataset $D$ cropped from the original image $\mathcal{D}$. Each patch is represented as a vertex $\upsilon \in V$. Two vertices $\upsilon$ and $\mu$ of $V$ such that $(\upsilon, \mu) \in E$ are called to be adjacent. Let $\upsilon = (x_i, y_i)$ and $\mu = (x_{i+1}, y_{i+1})$, we denote all possible adjacent relationships between $\upsilon$ and $\mu$ as: (1) horizontal:  $|x_{i+1} - x_{i}| = p$; (2) vertical: $|y_{i+1} - y_{i}| = p$; and (3) diagonal: $\sqrt{{(x_{i+1} - x_{i})}^2 + {(y_{i+1} - y_{i})}^2} = \sqrt{2} \times p$. With these characteristics, the positions of the patches can be identified in the original image. 
The graph-coloring-based splitting procedure is written in pseudo-code from step 1 to step 4.


Based on the assumption, we study different split strategies to scramble these identifications and reduce the risk of restoring the original dataset from the image fragments by the adversary. On the one hand, we adopt the graph-coloring-based split strategies~\cite{matula1983smallest, kubale2004graph,deo2006discrete}, including `largest\_first', `random\_sequential', `smallest\_last', `independent\_set', `connected\_sequential', `saturation\_largest\_first', 
to split the entire dataset $D$ into different sub-datasets $d_{p,1}, ..., d_{p, N}$, such that no two adjacent vertices share the same color or dataset. On the other hand, we introduce a random data perturbation to preserve the sensitive coordinates on split datasets' labels by inserting random noise.  

We extract $(x, y)_\text{coord}$ as a data matrix $A_p$ of size ($a \times b$), $a<b$, from $d_{p,k}\subset D$. After normalization, we compute the covariance matrix of the normalized matrix $A_{p,c}$, and then computed the eigenvalues $\lambda$ and eigenvectors $\vec{V}$ so that we can get the top-k eigenvectors $\vec{V}_k$ to calculate $A_e$. Moreover, we transform the data into a new coordinate system and encrypt it into datasets $\{d_{e,1}, d_{e,2},..., d_{e, N}\}$. 
The pseudo-code is illustrated in step 5 to step 15. 
From the perturbed data, since we know the noise variance, we obtain the estimate coordinates $\hat{Y}$ from decryption by inversely transforming the eigenvector matrix $\vec{V}_k$ and $A_e$, i.e., $\hat{Y}=A_e \cdot \vec{V}_k^T$. 
Note that the size of the transformed data matrix $A_e$ might differ from that of the original data matrix $A_p$. To address this discrepancy, we introduce random values $r$ to compensate for the size difference. Consequently, we can obtain a set of split image datasets with encrypted labels $\{d_{e,1}, ..., d_{e,k}, ..., d_{e,N}\}$. Besides, since we know the mappings of original labels and their corresponding encrypted labels, it is easy to measure the output utility shown in Eq.~\ref{eq:utility}. Furthermore, we calculate the average minimal privacy risk over the distributed datasets by Eq.~\ref{eq:rho_i_min}.
Empirical evidence indicates that the computation of eigenvalues and eigenvectors remains lightweight, even with up to 10,000 patches. 

\subsection{Pareto Trade-off Solution among Privacy, Cost, and Time}
The number of split datasets $\mathcal{N}$ is directly related to the number of cloud instances that need to be rented. Given split data $\{d_{e,1}, d_{e,2}, d_{e,N}\}$, CNN inference workload $\Theta$, hybrid cloud instances $M$, and user's budget $B$, our PriCE establishes a decision-making process based on MOO for resource planning.

The MOO problem is a classical integer programming problem since the variables $\vec{x}$ are restricted to be integers. We first find available optimal solutions of the bi-objective optimization problem that minimizes $f_2$ (Eq.~\ref{eq:f2}) and $f_3$ (Eq.~\ref{eq:f3}) under the \textit{budget}, and together with all available split strategies that generate $f_1$ (Eq.~\ref{eq:f1}), obtaining all feasible solutions as $\vec{X}$. Due to the trade-offs among minimizing lower bound on privacy risk $f_1$, monetary cost reduction $f_2$, and makespan minimization $f_3$, we then seek Pareto trade-off solutions $\vec{x}^*\in \vec{X}$, where no solution is superior to another in all objectives (See section~\ref{sec: problem_formulation}).
Pareto front is the set of all such non-dominated Pareto optimal solutions. From the mathematical point of view, the definition of the dominance between two candidate solutions can be expressed as $\vec{x_1}$ dominates $\vec{x_2}$ if $f_i(\vec{x_1}) \leq f_i(\vec{x_2})$, $\forall i=1,2,..., n$. Therefore, a solution $\vec{x}^*\in \vec{X}$ is called to be nondominated or Pareto optimal if and only if there does not exist any other point $\vec{x}\in \vec{X}$, such that $\vec{F}(\vec{x}) \leq \vec{F}(\vec{x^*})$ and $f_i(\vec{x})<f_i(\vec{x^*})$ for at least one function~\cite{marler2004survey}, in which $\vec{F}(\vec{x})= [f_1(\vec{x}), f_2(\vec{x}), f_3(\vec{x})]$. The MOO problem-solving procedure is written in the steps 16 to 18, as shown in Algorithm~\ref{alg: privacy method}.

\section{Experiments and Evaluation}\label{sec:experiments}
This section details the experimental setup, demonstrates the visualized results, and evaluates the outcomes for validation. We implemented our algorithm PriCE in Python and evaluated the capability and quality of response of our algorithm via simulation when used to answer the questions in Section~\ref{sec:intro}. It is important to note that the results presented here are, in whole or in part, based on data generated by the TCGA Research Network: \url{https://www.cancer.gov/tcga}. The original data and experimental results are available online\footnote{The source code is available online. \url{https://github.com/yuandou168/PriCE}}. 

\subsection{Experimental Setup}

We conducted extensive experiments on a dedicated remote server equipped with 6 cores/12 threads@3.6GHz, 64GB DDR4 RAM, and 2x512 GB NVMe SSD and a private GPU server equipped with a Tesla T4 16GB device. To collect benchmark data, we used a real-world CNN ensemble application for artifact detection developed by Kanwal \textit{et al.}~\cite{kanwal2024equipping} as our CNN workloads. Specifically, it is the ensemble of five CNNs. 
The complexity of the inference workloads has been measured by the total parameters, FLOPs, batch size, and memory usage. 

For the evaluation of solutions, we have investigated 25 commercial GPU servers offered by Fluidstack\footnote{Cloud GPU servers. \url{https://console2.fluidstack.io/virtual-machines}} and 2 private GPU servers offered by universities, located in different cities across the Netherlands, Norway, USA, Iceland, and India. Since these cloud instances have different configurations, we investigated their relative performance based on TPU review data about GPUs\footnote{GPU Database--TechPowerUp. \url{https://www.techpowerup.com/gpu-specs/}} and collected the workload performance over the private T4 GPU server offered by the university of Amsterdam. Besides, since the bandwidths are various with different geo-locations, we refer to fixed upload Internet speeds provided by SpeedTest\footnote{Network Speed Test.\url{https://www.speedtest.net/performance}} to simplify the network environments. More technical details have been presented in the source code.

\subsection{Visualization}
\begin{figure}[!htb]
    \centering
    \includegraphics[width=0.98\textwidth]{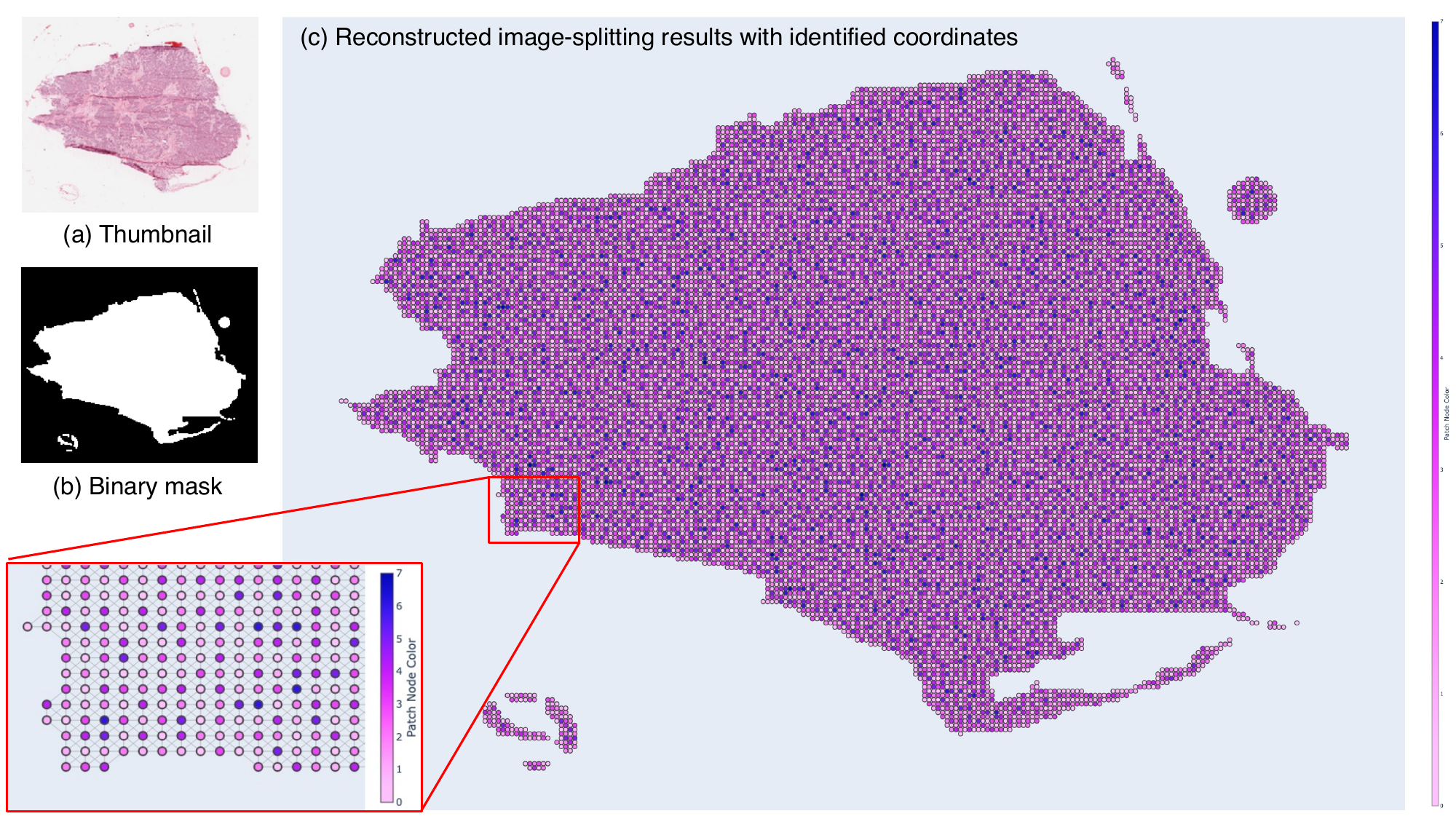}
    \caption{Visualization of the image-splitting: (a) the thumbnail picture of the original medical image, (b) the binary mask picture, and (c) the reconstructed graph from estimated coordinates after decryption.}
    \label{fig:visual1}
\end{figure}

In this study, we utilized a WSI named `TCGA-E9-A1N3-01Z-00-DX1' for demonstration, which is available on the TCGA repository\footnote{\url{https://portal.gdc.cancer.gov/image-viewer/MultipleImageViewerPage?caseId=03c143e0-d8a1-4d60-a4a3-df0501fc6b6e}}. In Fig.~\ref{fig:visual1}, we demonstrate that PriCE can effectively split a large medical image into different sub-datasets using a graph-coloring-based strategy, ensuring that no two adjacent patches are placed in the same sub-dataset. 
 
Fig.~\ref{fig:visual1}a represents the thumbnail picture of the original WSI and Fig.~\ref{fig:visual1}b is the binary mask pictures after removing the background of the large image. The patch nodes from the original image represent a set of colored images with the size of 224 $\times$ 224 pixels; each of them has a unique label that contains the coordinate information of the image to identify its position in the original image. First, we can see that PriCE can perform graph-coloring-based splitting with the coloring assignment to distribute image samples for the WSI. For example, we obtain eight sub-datasets with the `random sequential' graph-coloring strategy. Then, we adopt the data perturbation method to the split patch labels to hide the sensitive coordinate information; meanwhile, we examine the output utility that measures how close the estimated label of the privacy-preserving algorithm is to its desired output. In Fig.~\ref{fig:visual1}c, we plot the reconstructed graph with identified coordinates after the decryption procedure. 

For more complex and diverse WSIs, we can apply our method to extract various graphs by adding patch nodes and edges, choose, and run different strategies to generate different split datasets. Practically, we tested our PriCE for a number of TCGA WSIs. The outcomes always match the original coordinates, achieving the perfect output utility. 

\subsection{Reduction of the Average Lower Bound on Privacy Risk}
We bring a unique perspective by introducing our novel PriCE algorithm with `graph-coloring-based split' strategies (i.e., $\mathcal{S}_{graph}$), which we compare with the commonly used `average split with shuffle or without shuffle' ones (i.e., $\mathcal{S}_{avg}$).  
Table~\ref{tab:comparedprivacy} presents the overall comparisons of the number of split dataset and the average minimal privacy risk scores of the image-splitting strategies. Using graph-based split strategies, we achieved the minimum number $\mathcal{N}$ of split datasets of varying sizes. However, the average split strategy is not able to obtain that number except for customizing it by users. To make the comparisons fair, we custom the same $\mathcal{N}$ when splitting the datasets in the average cases. 
\begin{table}[!htb]
\scriptsize
    \centering
    \caption{Comparisons of split strategies, the number of split datasets, and the average minimal privacy risk scores. 
    }\label{tab:comparedprivacy}{\renewcommand{\arraystretch}{1.6}
    \resizebox{0.98\textwidth}{!}{
    \begin{tabular}{|c|c|c|c|c|c|}
    \hline
       {} & \multirow{2}*{\textbf{ {Strategy}}} & \multirow{2}*{ {$\mathcal{N}$}} & \multicolumn{3}{c|}{ {$\bar{\rho}_{min}$}}  \\ \cline{4-6} 
        {} & {} & {} &  {$x_{coord}$} &  {$y_{coord}$} &  {$\sum (x, y)$}\\ \hline
        \multirow{6}{*}{\rotatebox{90}{\textbf{{Graph-based Split}}}} 
         & \cellcolor{green!15} {saturation\_largest\_first} &  \cellcolor{green!15} {4} &  \cellcolor{green!15} {0.1835$\pm$0.001} &  \cellcolor{green!15} {0.1389$\pm$0.0007} &  \cellcolor{green!15} {0.3224} \\ 
        \cline{2-6}
        
        & \cellcolor{blue!25} {smallest\_last} & \cellcolor{blue!25} {5} & \cellcolor{blue!25} {0.1523$\pm$0.0564} & \cellcolor{blue!25} {0.1169$\pm$0.0409} & \cellcolor{blue!25} {0.2692} \\ \cline{2-6} 
        
        & \cellcolor{LightCyan} {connected\_sequential} & \cellcolor{LightCyan} {6} & \cellcolor{LightCyan} {0.1281$\pm$0.0738} & \cellcolor{LightCyan} {0.0987$\pm$0.055} & \cellcolor{LightCyan} {0.2268}\\ \cline{2-6}
        
        & \cellcolor{red!15} {independent\_set} & \cellcolor{red!15} {7} & \cellcolor{red!15} {0.1108$\pm$0.0798} & \cellcolor{red!15} {0.0859$\pm$0.0597} & \cellcolor{red!15} {0.1967}  \\ \cline{2-6}
        
        &\cellcolor{yellow} {largest\_first} & \cellcolor{yellow} {8} & \cellcolor{yellow} {0.1101$\pm$0.058} & \cellcolor{yellow} {0.0879$\pm$0.0445} & \cellcolor{yellow} {0.198} \\ \cline{2-6}
        
        & \cellcolor{yellow} {random\_sequential} & \cellcolor{yellow} {8} & \cellcolor{yellow} {0.1106$\pm$0.0571} & \cellcolor{yellow} {0.0881$\pm$0.0437} & \cellcolor{yellow} {0.1987} \\ \cline{2-6}
        \hline
        
        \multirow{10}{*}{\rotatebox[origin=c]{90}{\textbf{ {Average Split w/wo Shuffle}}}} &  \cellcolor{green!15} {is\_shuffled\_True} &  \cellcolor{green!15} {4} &  \cellcolor{green!15} {0.1832$\pm$0.0004} &  \cellcolor{green!15} {0.1386$\pm$0.0005} &  \cellcolor{green!15} {0.3218} \\ 
        \cline{2-6}
        
        & \cellcolor{blue!25} {is\_shuffled\_True} & \cellcolor{blue!25} {5} & \cellcolor{blue!25} {0.1613$\pm$0.0003} & \cellcolor{blue!25} {0.1255$\pm$0.0002} & \cellcolor{blue!25} {0.2868} \\ \cline{2-6}
        
        & \cellcolor{LightCyan} {is\_shuffled\_True} & \cellcolor{LightCyan} {6} & \cellcolor{LightCyan} {0.1448$\pm$0.0005} & \cellcolor{LightCyan} {0.1145$\pm$0.0005} & \cellcolor{LightCyan} {0.2593} \\ \cline{2-6}
        
        & \cellcolor{red!15} {is\_shuffled\_True} & \cellcolor{red!15} {7} & \cellcolor{red!15} {0.132$\pm$0.0003} & \cellcolor{red!15} {0.1061$\pm$0.0005} & \cellcolor{red!15} {0.2381} \\ \cline{2-6}
        
        & \cellcolor{yellow} {is\_shuffled\_True} & \cellcolor{yellow} {8} & \cellcolor{yellow} {0.1213$\pm$0.0007} & \cellcolor{yellow} {0.0989$\pm$0.0007} & \cellcolor{yellow} {0.2202} \\ \cline{2-6}
        
        & \cellcolor{green!15} {is\_shuffled\_False} &  \cellcolor{green!15} {4} &  \cellcolor{green!15} {0.1831$\pm$0.0009} &  \cellcolor{green!15} {0.1386$\pm$0.0002} & \cellcolor{green!15} {0.3217} \\ \cline{2-6}
        
        & \cellcolor{blue!25} {is\_shuffled\_False} & \cellcolor{blue!25} {5} & \cellcolor{blue!25} {0.1615$\pm$0.0011} & \cellcolor{blue!25} {0.1251$\pm$0.0008} & \cellcolor{blue!25} {0.2866} \\ \cline{2-6}
        
        & \cellcolor{LightCyan} {is\_shuffled\_False} & \cellcolor{LightCyan} {6} & \cellcolor{LightCyan} {0.1448$\pm$0.001} & \cellcolor{LightCyan} {0.1147$\pm$0.0006} & \cellcolor{LightCyan} {0.2595} \\ \cline{2-6}
        
        & \cellcolor{red!15} {is\_shuffled\_False} & \cellcolor{red!15} {7} & \cellcolor{red!15} {0.1318$\pm$0.0007} & \cellcolor{red!15} {0.106$\pm$0.0006} & \cellcolor{red!15} {0.2378} \\ \cline{2-6}
        
        & \cellcolor{yellow} {is\_shuffled\_False} & \cellcolor{yellow} {8} & \cellcolor{yellow} {0.1213$\pm$0.0005} & \cellcolor{yellow} {0.0985$\pm$0.0007} & \cellcolor{yellow} {0.2198} \\ 
        \cline{2-6}
    \hline
    \end{tabular}
    }
    }
\end{table}

The results show that (1) in the average split cases for the same number $\mathcal{N}$,  the minimal privacy risk scores with shuffle (`is\_shuffled\_True') and without shuffle (`is\_shuffle\_False') are very close to each other; (2) when the number of split datasets is four, the privacy risk scores of all cases are almost the same, though the one with `saturation\_largest\_first' is higher than the others; (3) as the number of datasets increases, the corresponding average minimal privacy risk score decreases; and (4) for the same number of split datasets, the average minimal privacy risk obtained by most graph-coloring-based split methods are generally lower but accompanied by a larger standard deviation. When $\mathcal{N}=7$, the difference between graph and average-based strategies is over 0.04, which is higher than when $\mathcal{N}=$ 5, 6, or 8, where the differences are below $0.02$, $0.04$, and $0.03$, respectively. As a result, the number of datasets $\mathcal{N}$ and corresponding split strategies significantly affect the privacy risk.

\subsection{Evaluation by Simulations}
\begin{figure}[!hb]
    \centering 
     \begin{subfigure}[b]{0.48\textwidth}
         \centering
         \includegraphics[width=\textwidth]{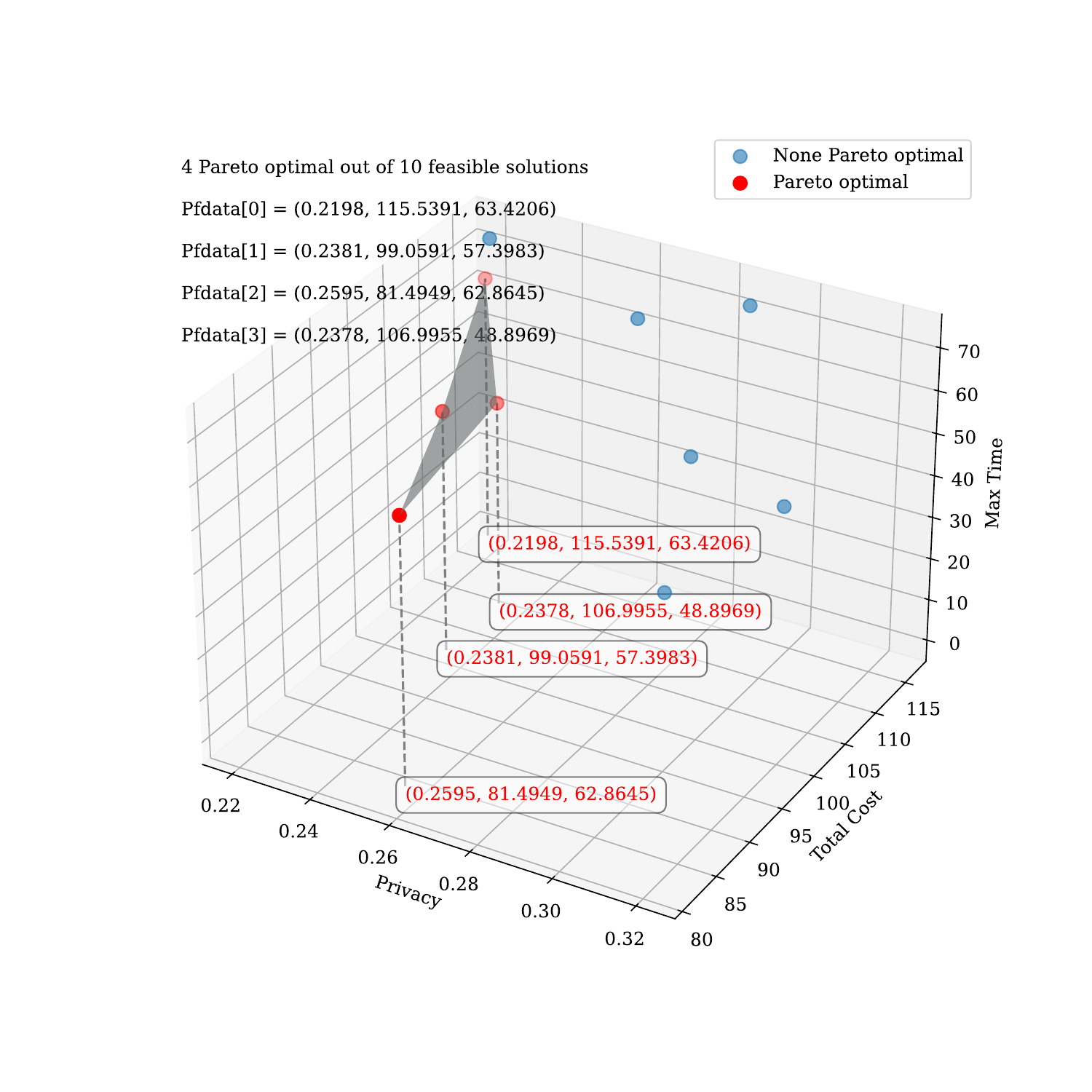}
         \caption{budget=120, $s\in \mathcal{S}_{\text{avg}}$.}
         \label{subfig:pareto120_onlyeven}
     \end{subfigure}
     \hfill
     \begin{subfigure}[b]{0.48\textwidth}
         \centering
         \includegraphics[width=\textwidth]{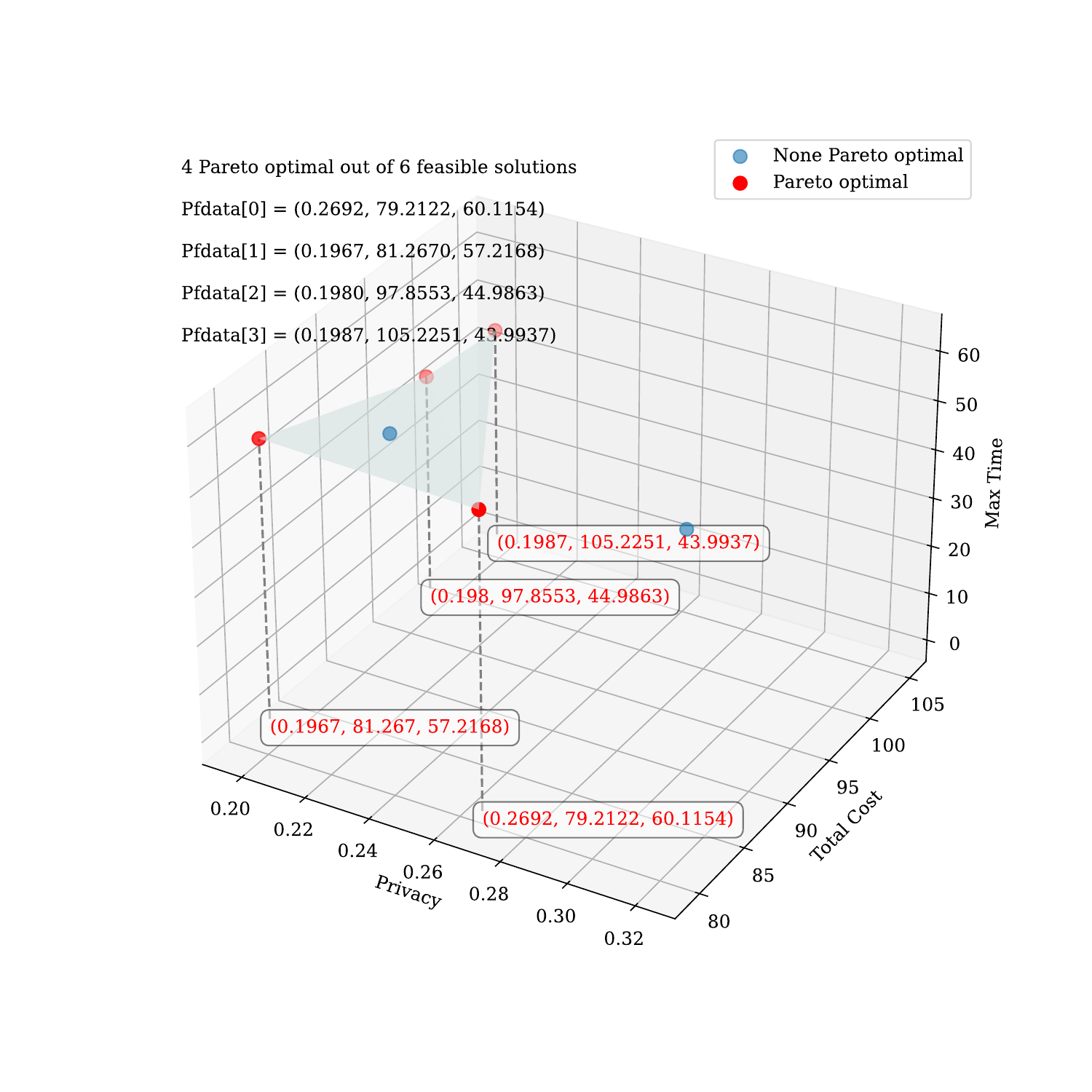}
         \caption{budget=120, $s\in \mathcal{S}_{\text{graph}}$.}
         \label{subfig:pareto120_onlygraph}
     \end{subfigure}
     \hfill
     \begin{subfigure}[b]{0.48\textwidth}
         \centering
         \includegraphics[width=\textwidth]{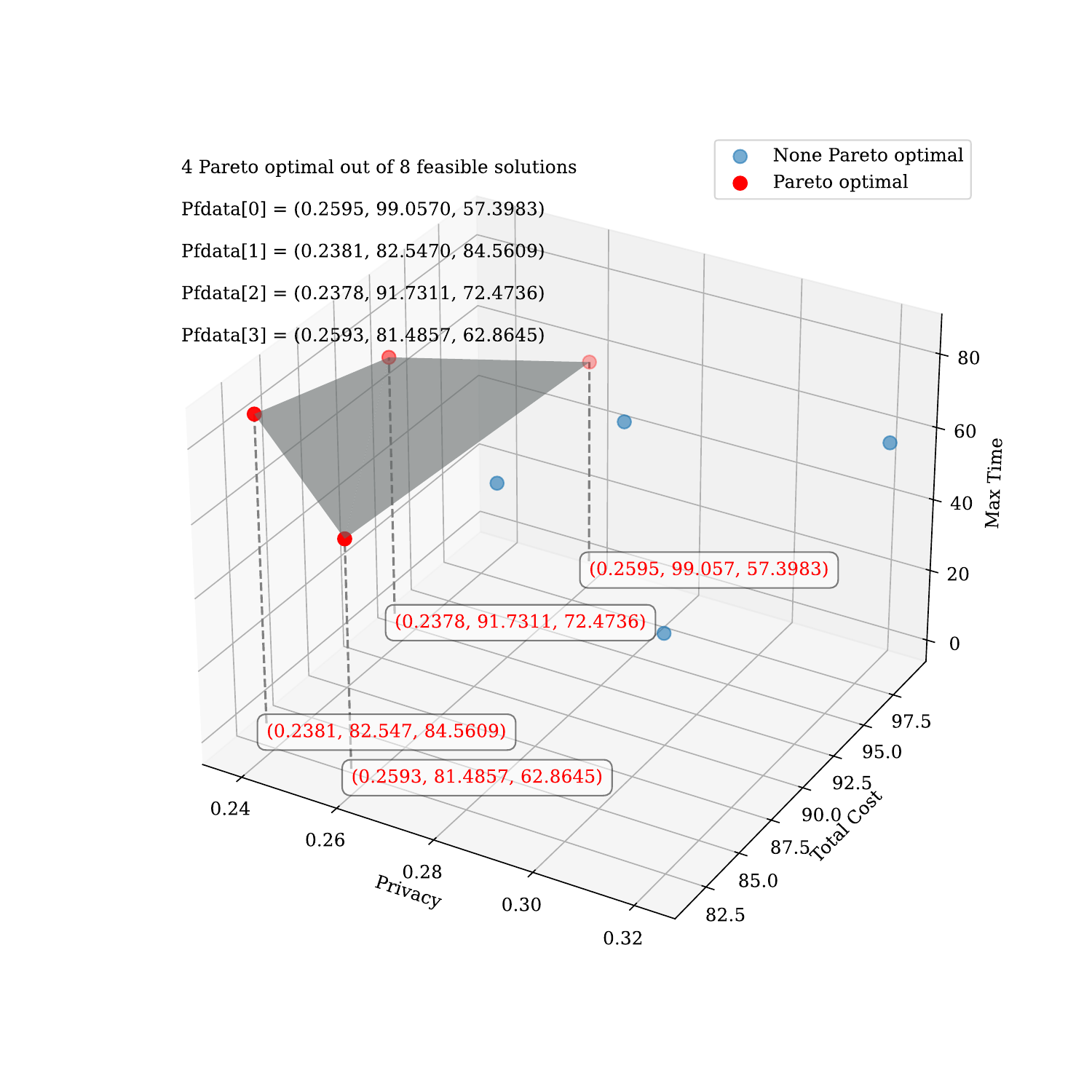}
         \caption{budget=100, $s\in \mathcal{S}_{\text{avg}}$.}
         \label{subfig:pareto100_onlyeven}
     \end{subfigure}
     \hfill
     \begin{subfigure}[b]{0.48\textwidth}
         \centering
         \includegraphics[width=\textwidth]{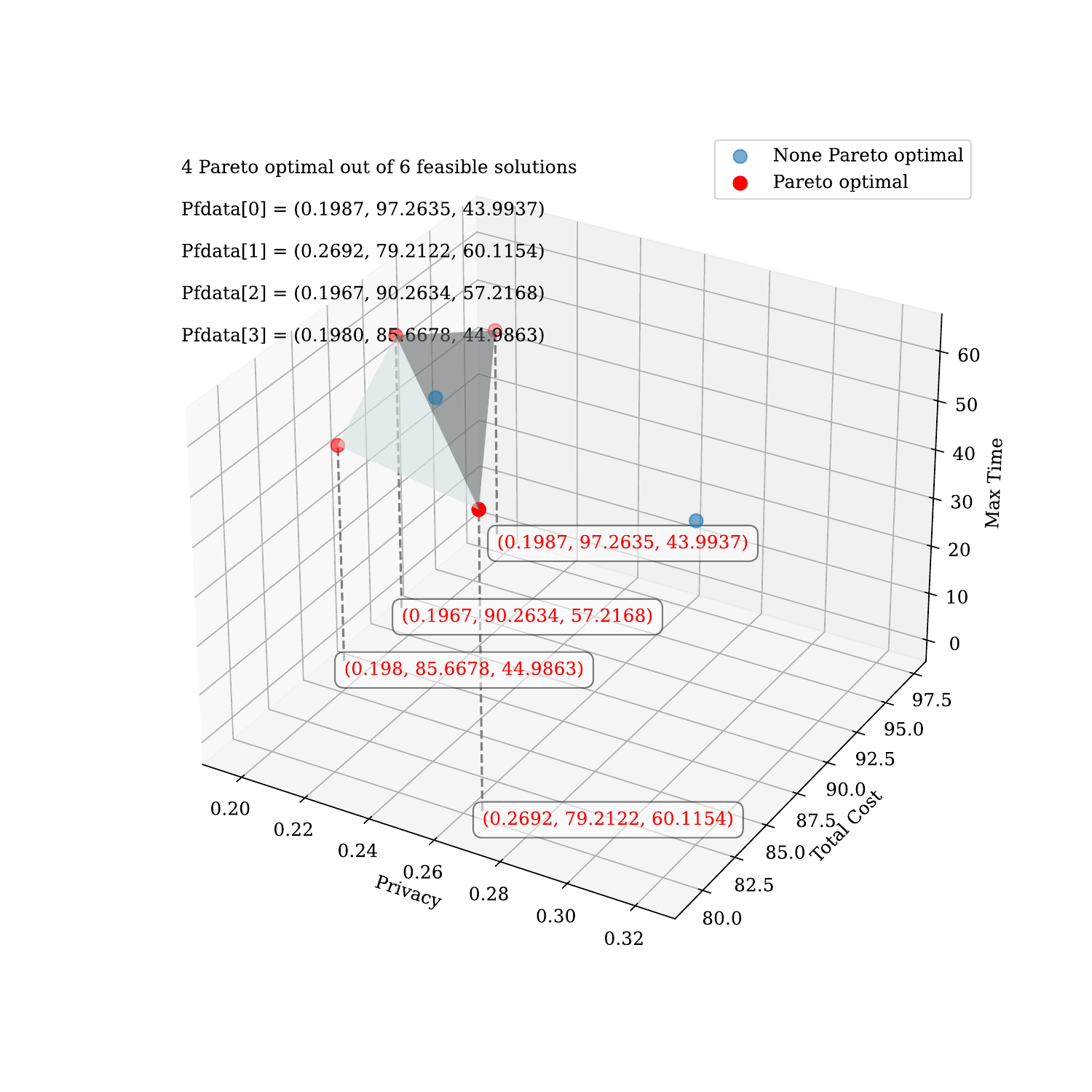}
         \caption{budget=100, $s\in \mathcal{S}_{\text{graph}}$.}
         \label{subfig:pareto100_onlygraph}
     \end{subfigure}

    \caption{The 3D Pareto trade-off solutions are compared as follows: (a) and (c) solutions fall within the only average split strategies \(\mathcal{S}_{\text{avg}}\), while (b) and (d) solutions are derived from the only graph-coloring-based split strategies $\mathcal{S}_{\text{graph}}$.
    }\label{fig:pareto}
\end{figure}

\begin{figure}[!ht]
    \centering 
     \begin{subfigure}[b]{0.48\textwidth}
         \centering
         \includegraphics[width=\textwidth]{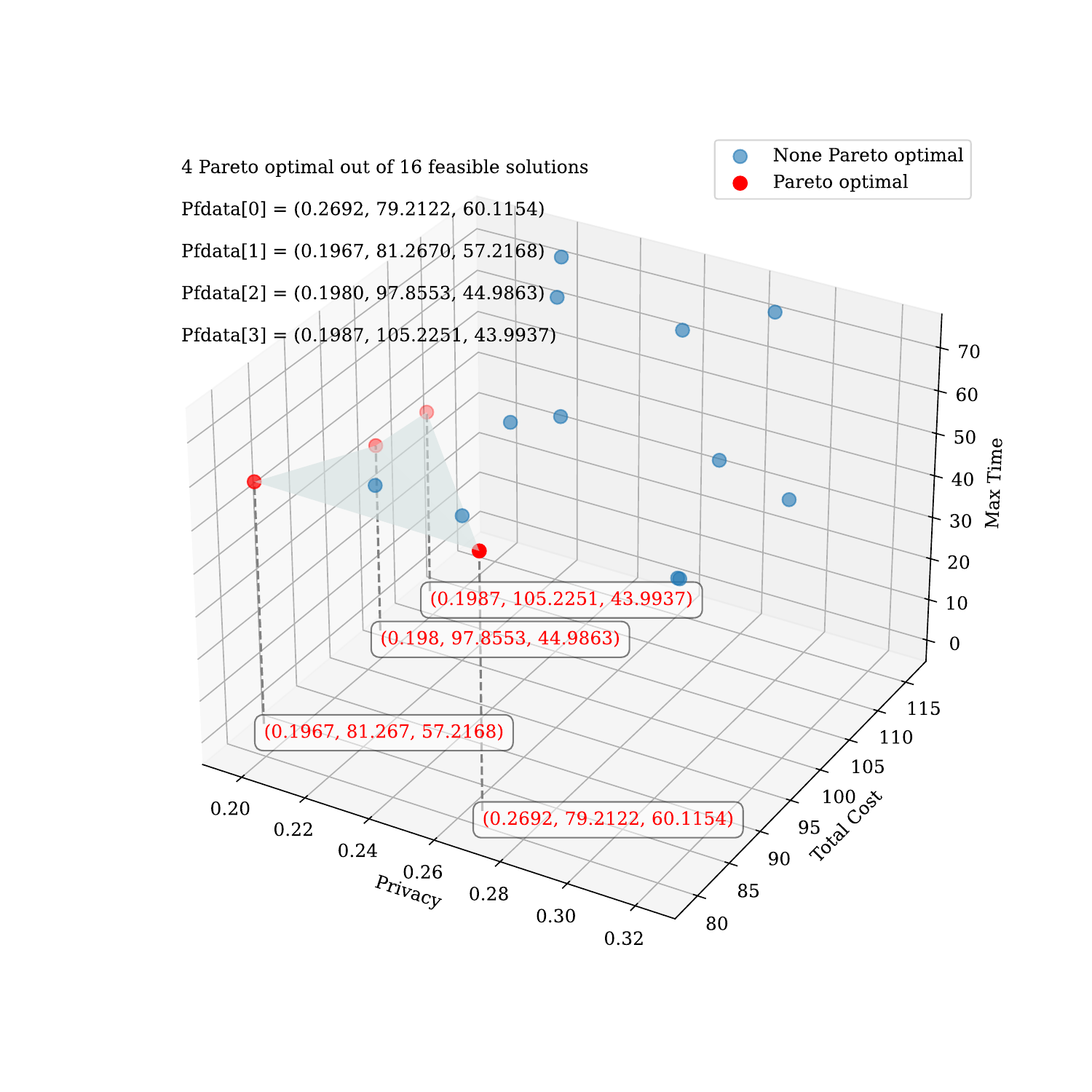}
         \caption{budget=120, $s\in \mathcal{S}_{\text{all}}$.}
         \label{subfig:pareto120}
     \end{subfigure}
     \hfill
     \begin{subfigure}[b]{0.48\textwidth}
         \centering
         \includegraphics[width=\textwidth]{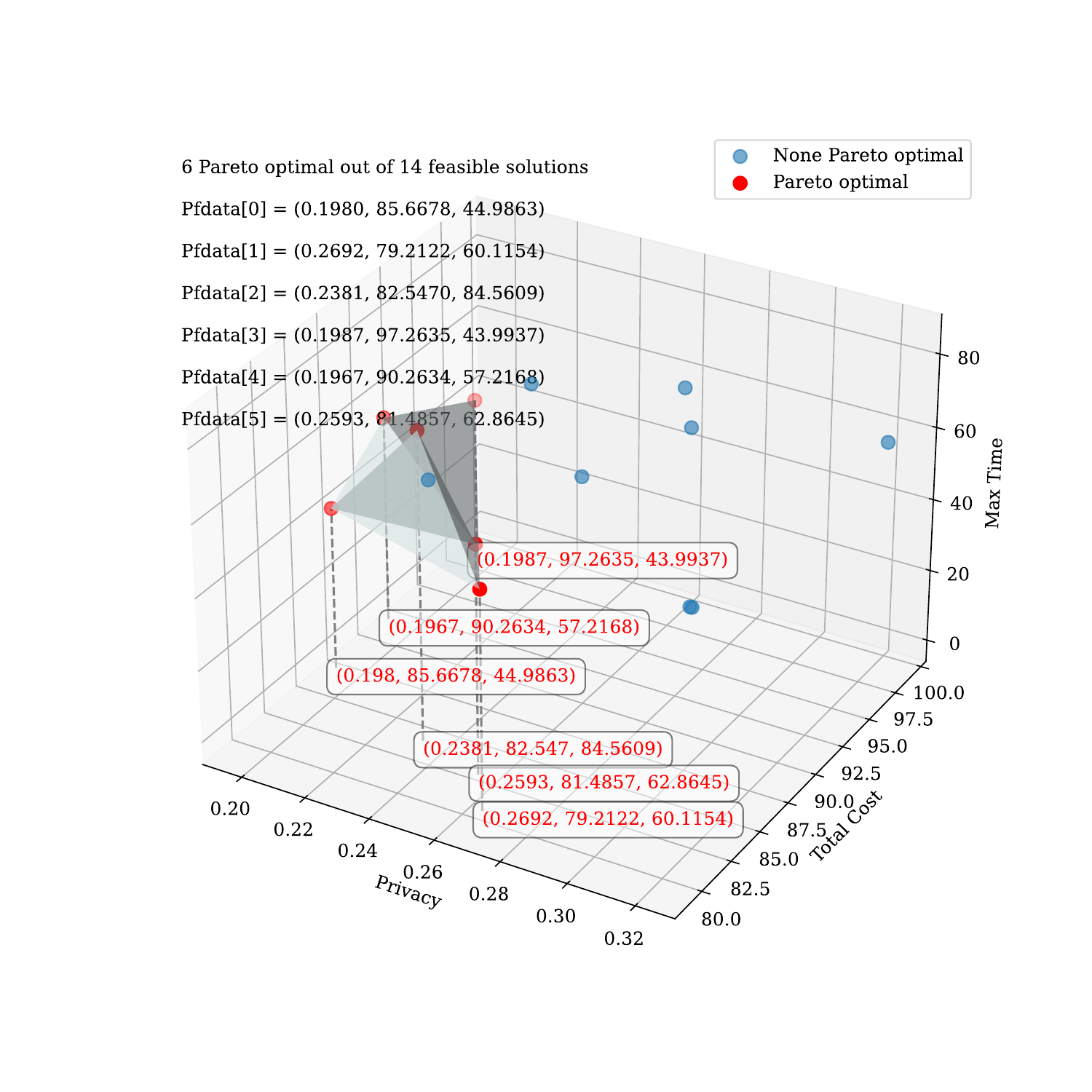}
         \caption{budget=100, $s\in \mathcal{S}_{\text{all}}$.}
         \label{subfig:pareto100}
     \end{subfigure}  

    \caption{Comparisons of the 3D Pareto trade-off solutions with all combined strategies $\mathcal{S}_{\text{all}}$ under two budget constraints: (a) budget = 120, and (b) budget = 100. 
    }\label{fig:paretoall}
\end{figure}

For Pareto optimal resource planning, we successfully obtained the Pareto trade-off solutions out of all feasible solutions from various split strategies under budget constraints through simulation experiments. Figs.~\ref{fig:pareto} and \ref{fig:paretoall} depict the Pareto trade-off solutions selected using the 3D Pareto front, comparing them across graph-based and average-based split strategies and two budget constraints.

When the budget is limited to $120$, we have identified four Pareto optimal solutions out of ten feasible solutions (4/10) in Fig.~\ref{subfig:pareto120_onlyeven}, within the scenario using only average split strategies ($s\in \mathcal{S}_{avg}$). In contrast, we have found four out of six Pareto trade-off solutions (4/6) within the scenarios using only graph-based split strategies ($s\in \mathcal{S}_{graph}$), as shown in Fig.~\ref{subfig:pareto120_onlygraph}. When decreasing the \textit{budget} to $100$, there are eight feasible solutions in the only average split case ($s\in \mathcal{S}_{avg}$), compared to ten solutions under the \textit{budget=120}, as seen in Fig.~\ref{subfig:pareto100_onlyeven}. This reduction is attributed to the absence of bi-objective optimal solutions for $f_2$ and $f_3$ identified by the problem solver when the number of average split datasets is eight. Remarkably, for the graph-coloring-based split, our PriCE method could still identify all Pareto trade-off resource planning solutions under the \textit{budget=100}, as depicted in Fig.~\ref{subfig:pareto100_onlygraph}. 

When combining both average and graph-based strategies ($s\in \mathcal{S}_{all}$), there are four Pareto trade-off solutions out of sixteen (4/16) when the budget is constrained by 120, as illustrated in Fig.~\ref{subfig:pareto120}. Notably, when compared to the separate cases, all four Pareto trade-off solutions (4/4) originate from the graph-coloring-based split strategies. Furthermore, when combining all feasible solutions from all split strategies and budget = 100, as depicted in Fig.~\ref{subfig:pareto100}, the majority of Pareto trade-off solutions (4/6) for the privacy-preserving and cost-effective problem are derived from the graph-coloring-based split strategies, with the remaining 2/6 solutions originating from the average split strategies. It can be inferred that when the budget becomes more constrained, privacy will be sacrificed.
We have observed when the split datasets are unbalanced, the resource allocation plan exhibits greater resilience in matching with heterogeneous cloud instances. This increased resilience arises from the enhanced diversity in execution times and costs, which maximizes the opportunities for efficient task assignments.

\section{Conclusion and Future Work}\label{sec:conclusion}
This paper investigates the workflow scheduling problem of privacy-preserving and cost-effective distributed inference using multiple GPU servers over hybrid clouds. We propose a novel solution, PriCE, which employs various image splitting strategies to enhance privacy and cost-efficiency. To the best of our knowledge, this is the first approach to address a privacy-aware scheduling problem that minimizes privacy risk while reducing makespan and cost within a budget in a privacy-preserving distributed system.
We conducted a comprehensive experimental evaluation using a real-world application for medical image artifact detection.
The results demonstrate that PriCE successfully takes the large number of patches from a gigapixel image and splits them using multiple graph-coloring-based strategies, yielding the desired output utility while lowering privacy risk, makespan, and monetary cost under the user's budget.
Further improvements in implementation might include making better use of secure distributed messaging or secrets handling across multiple clouds with high-level automated operations. Additionally, exploring the trade-offs between privacy overhead versus time and monetary cost in more complex scenarios could provide valuable insights.

\

\noindent {\textbf{Acknowledgment.}} We thank Mr. Zongxiong Chen for discussing the methodology and Mr. Aditya Shankar for reviewing the manuscript. This work has been partially funded by the European Union’s Horizon research and innovation program by CLARIFY (860627), BlueCloud-2026 (101094227), ENVRI-Hub Next (101131141), OSCARS (101129751), EVERSE (101129744), BioDT (101057437, via LifeWatch ERIC), by the LifeWatch ERIC and by the Dutch NWO LTER-LIFE project. 



\printbibliography

\end{document}